\documentclass[aps,english,floatfix,longbibliography]{revtex4-2}
\usepackage{natbib}
\usepackage{graphicx}
\usepackage{dcolumn}
\usepackage{bm}
\usepackage[
        colorlinks=true,
        urlcolor=black,
        citecolor=blue,
        linkcolor=blue]{hyperref}
\usepackage{siunitx}
\usepackage{amsmath}
\usepackage{amssymb}
\usepackage[normalem]{ulem}

\newcommand{\defeq}{\overset{\mathrm{def}}{=\joinrel=}}

\begin{document}

\preprint{APS/123-QED}

\title{Complex dynamical regimes of the Tayler-Spruit dynamo}

\author{Paul Barr\`{e}re$^{1,2}$}
\email{paul.barrere@unige.ch}
\author{J\'{e}r\^{o}me Guilet$^{2}$}
\email{jerome.guilet@cea.fr}
\author{Basile Gallet$^{3}$}
  \email{basile.gallet@cea.fr}
\author{Rapha\"{e}l Raynaud$^{4}$}
 \email{raphael.raynaud@cea.fr}

\affiliation{
$^{1}$Observatoire de Genève, Université de Genève, 51 Ch. Pegasi, 1290 Versoix,  Switzerland\\
$^{2}$Universit\'e Paris-Saclay, Universit\'e Paris Cit\'e, CEA, CNRS, AIM, 91191, Gif-sur-Yvette, France\\
$^{3}$Universit\'e Paris-Saclay, CNRS, CEA, Service de Physique de l'Etat condens\'{e}, 91191, Gif-sur-Yvette, France\\
$^{4}$Universit\'e Paris Cit\'e, Universit\'e Paris-Saclay, CNRS, CEA, AIM, F-91191 Gif-sur-Yvette, France}

\begin{abstract}
Astrophysical dynamos feature various spatial structures and dynamical regimes, ranging from hemispherical magnetic fields to the random reversals of the geodynamo. The Tayler-Spruit dynamo, recently confirmed in direct numerical simulations, has been invoked to explain angular momentum transport in stellar radiative zones and magnetar formation in a proto-neutron star spun up by fallback accretion. Whether this dynamo mechanism can lead to different dynamical regimes remains an open question. Using three-dimensional direct numerical simulations, we model the dynamics of a stably stratified spherical Couette flow, with the outer sphere rotating faster than the inner one. While the generation of strong stationary and hemispherical dynamos has been observed in our previous studies, we report for the first time the existence of reversals and complex temporal dynamics. We observe that the dynamics is strongly correlated with the equatorial symmetry breaking of the flow. Focusing on a fiducial dynamo simulation, we propose an interpretation of its dynamics that consists in the coupling of two large-scale magnetic modes with opposite equatorial symmetries by the flow symmetry breaking. While this interpretation captures the simplest observed dynamics, the nonlinear interaction between a higher number of magnetic modes is certainly required to describe more complex regimes. The wide diversity of dynamical regimes generated by the Tayler-Spruit dynamo may have interesting implications for the geometry of the neutron star magnetic fields, and therefore neutron star emissions.
\end{abstract}

\maketitle

\section{Introduction}
Magnetic fields are present in most astrophysical objects, from neutron stars (NS) to galaxies, including stars, accretion disks and planets. They span a wide range of strengths, from $10^{-6}\,\unit{G}$ in hot intracluster medium~\citep{carilli2002,bonafede2010} to $10^{15}\,\unit{G}$~\citep{kaspi2017} in magnetars. The question of their amplification and sustainment is made challenging by the nonlinear and multi-scale nature of the problem and remains an area of active research (see e.g. the reviews~\citep{brandenburg2005a,kulsrud2008,federrath2016,rincon2019,tobias2021}). Most observed astrophysical dynamos can be sorted in the following categories: (i) the galactic helical dynamo, which may be driven by supernova (SN)-induced turbulence~\citep{brandenburg2015,subramanian2019}, (ii) the dynamo driven by rotating convection, which is applied to stellar convective zones~\citep{kapyla2023,charbonneau2023}, Earth outer core~\citep{landeau2022}, white dwarf stars~\citep{isern2017}, and proto-neutron stars (PNS)~\citep{Raynaud2020,raynaud2022}, (iii) and the subcritical dynamos driven by magnetohydrodynamic (MHD) instabilities in shear flows, which could operate in stably stratified media, such as accretion disks~\citep{hawley1996,rincon2007,lesur2008} and stellar radiative zones~\citep{spruit2002,cline2003,vasil2024}. 

The dynamo problem is all the more difficult that astrophysical magnetic fields show a rich variety of complex geometries and evolve in a very wide range of physical regimes. The magnetic fields can also harbour complex temporal dynamics, like the chaotic reversals and excursions of the geomagnetic field~\citep{korte2019} or the 22-year solar cycle~\citep{hathaway2010}. Similar dynamics were also observed in the vicinity of the dynamo threshold in the Von Karman Sodium (VKS) experiment, which displayed a self-sustained magnetic field in a turbulent Von Karman swirling flow of liquid sodium forced by two counter-rotating bladed disks~\cite{monchaux2007,berhanu2007,ravelet2008,monchaux2009}.

Numerical simulations constitute a crucial tool to study these magnetic states. Several numerical models of rotating, convection-driven dynamo manage to reproduce hemispherical~\cite{stanley2008,amit2011,dietrich2013,landeau2017}, oscillatory magnetic fields~\cite{strugarek2017,strugarek2018}, or both~\cite{grote2000,raynaud2016}. Complex dynamics are also observed in numerical simulations of the magnetorotational instability (MRI)-driven dynamos in stably stratified shear flows~\cite{herault2011,riols2013,riols2017,reboul2021a,reboul2022}.

In this article, we focus on the \emph{Tayler-Spruit} dynamo, which is driven by an MHD instability called the Tayler instability. This instability feeds off a toroidal field in a stably stratified medium due to the presence of an electric current along the axis of symmetry~\cite{tayler1973,goossens1981}. The resulting electromotive force induces a large-scale poloidal magnetic field that gets sheared, thereby strengthening the toroidal component and closing the dynamo loop~\citep{spruit2002,denissenkov2007,zahn2007,fuller2019}. While this dynamo process remained elusive in numerical simulations, different analytical prescriptions were widely implemented in 1D stellar evolution codes and show promising results to explain helio/asteroseismic observations~\cite{cantiello2014,eggenberger2019b,eggenberger2019c,denhartog2020,griffiths2022,eggenberger2022,fuller2022}. The Tayler-Spruit mechanism in PNS spun up by supernova fallback accretion is also a promising scenario to explain magnetar formation, namely, NS whose high-energy emission is powered by the dissipation of extreme magnetic fields ($\SI{e14}{}-\SI{e15}{G}$)~\cite{barrere2022,barrere2023,barrere2024a,igoshev2025}.

Similarly to the MRI-driven dynamo, the transition to the Tayler-Spruit dynamo is subcritical, since it is observed for flows that are stable against infinitesimal magnetic perturbations and requires finite-amplitude disturbances to be triggered. This type of dynamo is made possible by the nonlinear nature of the magnetohydrodynamic (MHD) equations. This subcritical character complexifies their realisation in direct numerical simulations (DNS), which may explain why the Tayler-Spruit dynamo was identified for the first time in DNS only recently~\cite{petitdemange2023}. Our DNS of MHD Taylor-Couette flow also demonstrated the existence of two subcritical and bistable dynamo branches harbouring distinct magnetic field geometry and intensities~\cite{barrere2023}: (i) strong and equatorially symmetric, (ii) weak and hemispherical. In the following, we will refer to these branches as the \emph{strong} and the \emph{hemispherical} branches, respectively. For magnetic Prandtl numbers (ratio of the viscosity to the magnetic diffusivity) $1\leq Pm \leq 4$, we also showed that the strong branch can be maintained in a self-sustained state for stratifications with a ratio of the Brunt-Väisälä frequency to the rotation rate $N_o/\Omega\leqslant 4$, while it becomes transient for $4<N_o/\Omega\leqslant 10$~\cite{barrere2024a}. However, new DNS show that the strong dynamo can be maintained for $N_o/\Omega_o\lesssim130$ by using another setup with volumetric forcing for differential rotation instead of a Taylor-Couette configuration (i.e. boundaries rotating with fixed rates)~\citep{barrere2026}.

In this article, we report the discovery of complex temporal dynamics and transient magnetic field reversals produced by the Tayler-Spruit dynamo in the Tayler-Couette setup. After introducing the MHD equations and the numerical setup in Sect.~\ref{sec:num}, we describe the wide variety of dynamics we observe in our set of DNS in Sect~\ref{sec:param}. In Sect.~\ref{sec:cycle}, despite the high complexity and the diversity of observed dynamics, we propose an interpretation of the magnetic field dynamics operating in one DNS. Finally, we discuss the results and the astrophysical implications in Sect.~\ref{sec:conclu}.

\section{Numerical modelling}\label{sec:num}
\subsection{Governing equations}
Like in our previous numerical study~\citep{barrere2024a}, we consider a stably stratified and Boussinesq MHD flow evolving in a spherical Taylor-Couette configuration: the flow is confined between two concentric spheres of radius~$r_i$ and $r_o=4r_i$ defining the shell gap $d\defeq r_o -r_i=0.75r_o$. Both spheres rotate as a solid body with the respective rates $\Omega_i$ and $\Omega_o$, whose difference is parametrised by the Rossby number 
\begin{equation}\label{eq:rossby}
    Ro\defeq 1-\frac{\Omega_i}{\Omega_o}\in[0.75,1.4]\,.
\end{equation}
Note that the spheres are therefore in counter-rotation  when $Ro>1$ (and rotate in the same direction when $Ro<1$). We apply no-slip, fixed temperature, and electrically insulating conditions on both shells. We assume uniform kinematic viscosity $\nu$, thermal diffusivity $\kappa$, and magnetic diffusivity $\eta$, which are characterised by the thermal and magnetic Prandtl numbers
\begin{align}
    &Pr\defeq\frac{\nu}{\kappa}=0.1\,,\\
    &Pm\defeq\frac{\nu}{\eta}\in[1,4]\,,
\end{align}
respectively. In line with the Boussinesq approximation, which amounts to neglecting density variations everywhere except in the buoyancy term, we assume a uniform mass distribution, and hence a gravity that scales linearly with radius, $\mathbf{g} = -g_o r/r_o\mathbf{e}_r$, where $g_o$ denotes the gravitational acceleration at the outer sphere.

The governing equations are rendered dimensionless by scaling length in units of the shell gap $d$, time in units of the viscous time $d^2/\nu$, magnetic field in units of $\sqrt{\mu_0\rho\eta\Omega_o}$, where $\mu_0$ is the vacuum permeability, and temperature in units of temperature contrast between both spheres $\Delta T\defeq T_o-T_i$. The stable stratification is imposed by fixing $\Delta T>0$ and is parametrised in the equations by the Rayleigh number
\begin{equation}
    Ra\defeq \frac{d^4N_o^2}{\nu\kappa}=\frac{d^3\alpha g_o \Delta T}{\nu\kappa}\in[10^7,10^{11}]\,,
\end{equation}
where $N_o$ and $\alpha = \left.-{\partial \ln \rho}/{\partial T}\right|_{p}$ are the Brunt-Väisälä frequency at the outer boundary and the thermal expansion coefficient, respectively.
In the following, the thermal stratification will be characterised by $N_o/\Omega_o$, the ratio of the Brunt-Väisälä frequency to the rotation rate of the outer shell.
In the reference frame rotating with the surface at the angular velocity $\mathbf{\Omega}_o= \Omega_o \mathbf{e}_z$ characterised by the Ekman number
\begin{equation}\label{eq:ekman}
    E\defeq \frac{\nu}{d^2\Omega_o} =10^{-5}\,,
\end{equation}
the dimensionless Boussinesq MHD equations read
\begin{align} \label{eq:1}
    &\partial_t\mathbf{v} + (\mathbf{v}\cdot\nabla) \mathbf{v} + \frac{2}{E}\mathbf{e}_z\times\mathbf{v} = -\nabla p' + \frac{Ra}{Pr}T' \mathbf{e}_r 
         + \frac{1}{E\,Pm} (\nabla\times \mathbf{B})\times \mathbf{B} + \nabla^2\mathbf{v}\,, \\ \label{eq:3}
    &\partial_t T' + (\mathbf{v}\cdot\nabla) T' =\frac{1}{Pr}\nabla^2 T'\,,\\ \label{eq:4}
    &\partial_t\mathbf{B} =\nabla\times(\mathbf{v}\times\mathbf{B})+\frac{1}{Pm}\nabla^2\mathbf{B}\,,\\ \label{eq:5}
    &\nabla \cdot \mathbf{v} =0\,,\:\nabla\cdot\mathbf{B} =0\,, 
\end{align}
where $\mathbf{B}$ is the magnetic field, $\mathbf{v}$ is the velocity field, $p'$ is the reduced pressure (i.e. the pressure divided by the density), and $T'$ is the temperature perturbation from the stationary profile. $\mathbf{e}_z$ and $\mathbf{e}_r$ are the unit vectors of the axial and the spherical radial directions, respectively. Note that these equations ignore the effects of chemical composition and local heat sources.

\begin{figure*}
    \centering
    \includegraphics[width=\textwidth]{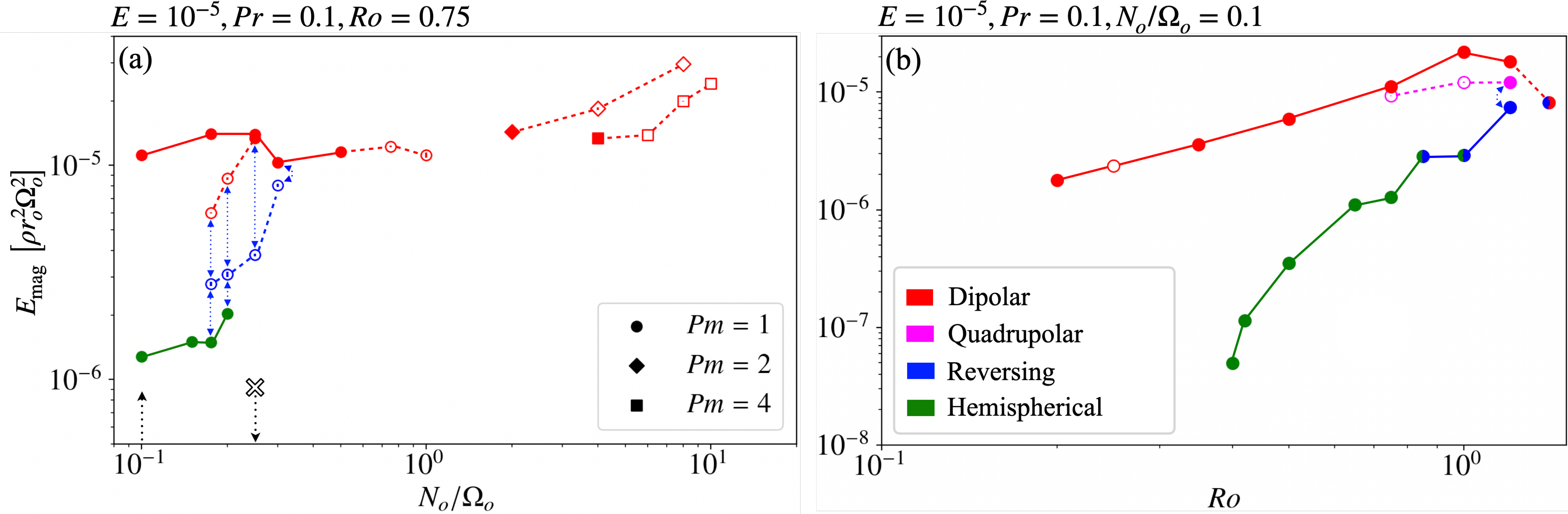}
    \caption{Bifurcation diagrams representing the time- and volume-averaged total magnetic energy as a function of the ratio of the Brunt-Väisälä frequency to the rotation rate of the outer sphere $N_o/\Omega_o$ (a) and as a function of the Rossby number $Ro$ (b). The different colours represent different saturated states of the magnetic field generated by the Tayler-Spruit dynamo: the strong stationary dipolar and quadrupolar (red and pink), hemispherical (green), and reversing (blue) magnetic fields. Filled markers represent self-sustained and transient dynamo states, while empty markers indicate that the dynamo state is transient, i.e. it eventually converges towards either another dynamo or a non-dynamo state. The non-dynamo case is represented by black crosses and the black dotted arrow pointing downward. The black dotted arrow at $N_o/\Omega_o=0.1$ in panel (a) indicates that the non-dynamo state is unstable. The dotted blue arrows indicate transitions between two states at different energies in the same time series. When the energies are similar, the transitions are represented by circles with two colours. While the full lines connect the self-sustained states, the dashed lines link the transient states.}
    \label{fig:bifurcation_NO}
\end{figure*}

\subsection{Numerical aspects}
To integrate Eqs.~\eqref{eq:1}--\eqref{eq:5}, we use the open-source pseudo-spectral code MagIC (commit 2266201a5) \citep{wicht2002,gastine2012,schaeffer2013,ishioka2018}. The velocity and magnetic fields are decomposed into poloidal and toroidal components:
\begin{align}\label{eq:v_decomp}
    \mathbf{v} &= \nabla \times \nabla \times (W\mathbf{e}_r) + \nabla\times(Z\mathbf{e}_r)\,,\\
    \label{eq:B_decomp}
    \mathbf{B} &= \nabla \times \nabla \times (g\mathbf{e}_r) + \nabla\times(h\mathbf{e}_r)\,,
\end{align}
where $W$ and $Z$ are the poloidal and toroidal kinetic potentials, respectively, while $g$ and $h$ are the magnetic ones. The horizontal (i.e. in colatitude $\theta$ and longitude $\phi$) and radial dependencies of these fields and reduced pressure $p'$ are then expanded into spherical harmonics and Chebyshev polynomials. For the time stepping, we use an implicit-explicit Runge-Kutta scheme defined in~\citet{boscarino2013}. 

The resolution is fixed at $(n_r,n_{\theta},n_{\phi})=(257,256,512)$ for all the runs presented in this article. This grid is refined enough to resolve with at least $10$ grid points the viscous and dissipation length scales defined by
\begin{equation}
    l_{\nu} \defeq \sqrt{\frac{2\nu E_{\rm kin}}{\Phi_{\nu}}}\,
\end{equation}
and
\begin{equation}
    l_{\eta} \defeq \sqrt{\frac{2\eta E_{\rm mag}}{\Phi_{\eta}}}\,,
\end{equation}
where $E_{\rm kin}$ and $E_{\rm mag}$ are the kinetic energy and magnetic energy, and $\Phi_{\nu}$ and $\Phi_{\eta}$ are the viscous and Ohmic dissipations. The resolution is also well suited to capture the turbulent flow produced by the Tayler-Spruit dynamo because at least 15 points resolve the characteristic turbulent length scale defined by
\begin{equation}\label{eq:l_turb}
    l_{\rm turb}\defeq \pi\frac{\sum_{\ell=0}^{\ell_{\rm max}}E_{\rm kin}^{m\neq0}(\ell)}{\sum_{\ell=0}^{\ell_{\rm max}}\ell E_{\rm kin}^{m\neq0}(\ell)}\,,
\end{equation}
where $\ell$ is the spherical harmonic degree. By starting from the runs called \emph{Ro0.75s} and \emph{Ro0.75w} from our previous article~\cite{barrere2023}, the numerical simulations are initialised using the nearby saturated state of a run with a weaker stratification. Using this procedure, the stratification is increased gradually to avoid losing the dynamo branch. 

\begin{figure*}
    \centering
    \includegraphics[width=\textwidth]{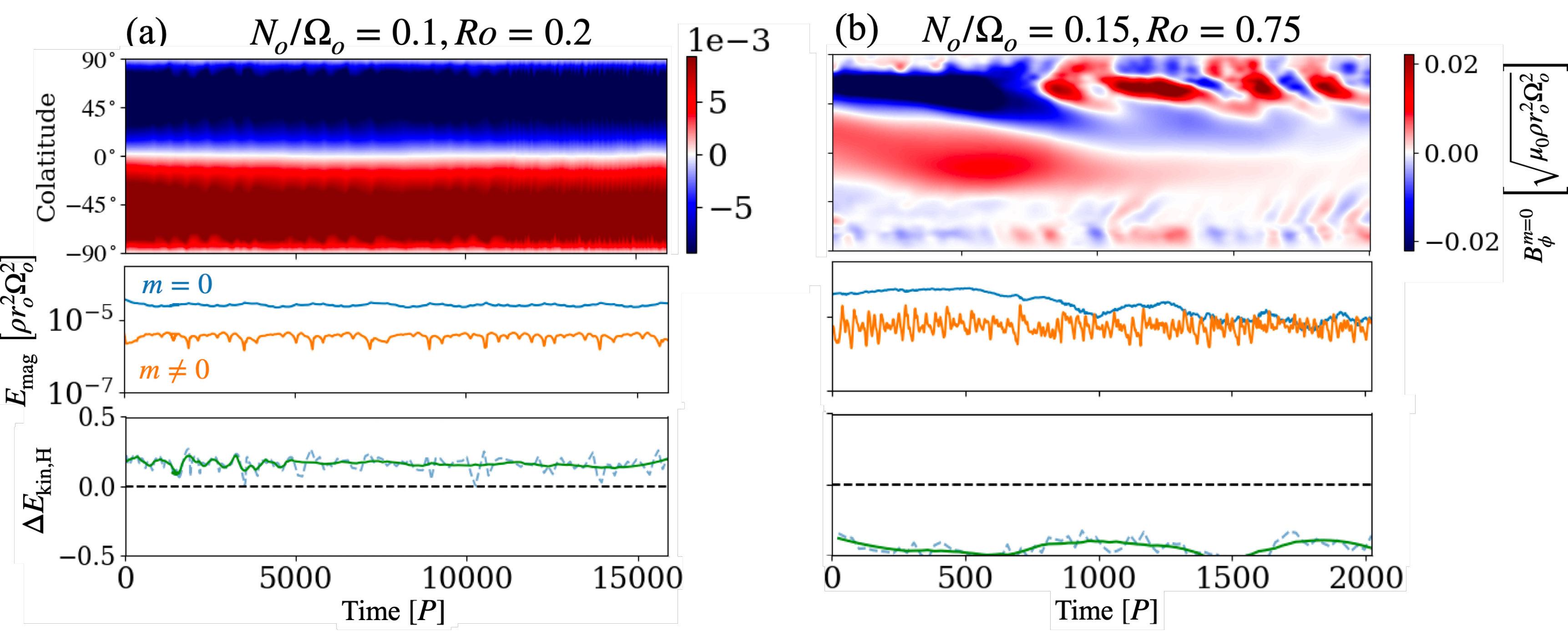}
    \caption{Butterfly diagram of the axisymmetric azimuthal magnetic field averaged over the radius interval $[0.42,0.5]\,r_o$ (top), time series of the axisymmetric and non-axisymmetric magnetic energies (middle), and time series of the parameter characterising the equatorial symmetry breaking of the flow $\Delta E_{\rm kin,H}$ (bottom) for the DNS at $N_o/\Omega_o=0.1,\,Ro=0.2$ (a) and $N_o/\Omega_o=0.15,\,Ro=0.75$ (b). While the blue dashed curve represents the calculated values of $\Delta E_{\rm kin,H}$, its smoothed version is plotted in green.}
    \label{fig:DipHem}
\end{figure*}

\subsection{Outputs}
While the shell gap and viscous time serve as units in MagIC, we present our results using the outer sphere radius $r_o$ and rotation period $P\defeq 2\pi/\Omega_o$ as length and time units, and we express energy densities in units of $\rho r_o^2 \Omega_o^2$. Hence, dimensionless quantity conversion reads
\begin{gather}
r_{\rm MagIC}^* \longrightarrow r^*=r_{\rm MagIC}^*\times\frac{r_o-r_i}{r_o}\:;\:t_{\rm MagIC}^* \longrightarrow t^* = t_{\rm MagIC}^* \times \frac{1}{2\pi E}\,,\\
E_{\rm MagIC}^* \longrightarrow E^*=  E_{\rm MagIC}^* \times\left(\frac{r_o-r_i}{r_o}\right)^2E^2\,,
\end{gather}
where $E$ is the Ekman number defined in Eq.~\eqref{eq:ekman} and must not be confused with an energy $E^*$.  Finally, we express the magnetic field strength in units of $\sqrt{\mu_0 \rho} r_o \Omega_o$, i.e. in terms of a Lorentz number defined as the ratio of Alfvén and rotational speeds. In bifurcation diagrams, energies are averaged over the time interval corresponding to the saturated dynamo state. In this article, we will also use other quantities to analyse our DNS: the electromotive force
\begin{equation}
    \mathbf{\mathcal{E}} \defeq \mathbf{v}^{m\neq0}\times\mathbf{B}^{m\neq0}\,,
\end{equation}
and the fraction of the $\Omega$-effect as defined by~\citet{schrinner2011}
\begin{equation}
    \left[\nabla\times (v_{\phi}\mathbf{e}_{\phi}\times \mathbf{B})\right]\cdot \mathbf{e}_{\phi}
\end{equation}
in the azimuthal component of the induction term
\begin{equation}
    \left[\nabla\times (\mathbf{v}\times \mathbf{B})\right]\cdot \mathbf{e}_{\phi}\,.
\end{equation}

To determine whether a dynamo state is self-sustained, we ran numerical simulations for at least two turbulent resistive times $t_{\rm turb,\eta}\defeq l_{\rm turb}^2/\eta\approx0.1 r_o^2/\eta$, where $l_{\rm turb}$ is the turbulent length scale defined in Eq.~\eqref{eq:l_turb}. For the simulations in which the dynamo effect is eventually lost, the potential transient dynamo state does not last more than one $t_{\rm turb,\eta}$. In contrast, the models with the parameters $Ro=0.75,N_o/\Omega_o\in\{0.2,0.25,0.3,0.5\}$ and $Ro=\{0.2,1.2\},N_o/\Omega_o=0.1$), which are either close to the dynamo threshold or showing several transitions between different states, have been run for at least one resistive time $t_{\eta}\defeq r_o^2/\eta\approx 10 \, t_{\rm turb,\eta}$.

The output data displayed in Figs.~\ref{fig:DipHem},~\ref{fig:butterfly}, and~\ref{fig:fiducial}(b,c) have been smoothed using the Savitzky–Golay filter implemented in the Python library \emph{Scipy} (function \emph{scipy.signal.savgol\_filter}). For the filter, we use a polynomial order of 2 and a filter window length of 11 samples, which represents a smoothing time interval of $\approx 280\,P$. The number of stored samples varies between $100$ and $700$ depending on the simulation integration time. This gives us a smoothed curve with the same trend as the original one and clarifies the system's trajectory in phase space.

Finally, we will often mention the symmetry of the magnetic field throughout the rest of the article. Like~\citet{knobloch1998}, we will consider a magnetic field to harbour a \emph{dipolar} symmetry when the poloidal magnetic potential $g$ is equatorially antisymmetric (i.e. the addition of the spherical harmonic degree and order $\ell+m$ is odd), while the toroidal potential $h$ is equatorially symmetric (i.e. $\ell+m$ is even). Conversely, a magnetic field has a \emph{quadrupolar} symmetry when $g$ and $h$ are equatorially symmetric and antisymmetric, respectively.

\section{Different spatial structures and temporal dynamics}\label{sec:param}
Figure~\ref{fig:bifurcation_NO} displays two bifurcation diagrams, in which all our set of DNS is plotted. When the stratification and the differential rotation are low, $N_o/\Omega_o\in[0.1,0.15]$ and $Ro\in[0.42,0.75]$, two dynamo states are reached subcritically and coexist in bistability: the strong stationary branch and the hemispherical one, which we already reported in a previous study~\citep{barrere2023}. Figure~\ref{fig:DipHem} shows the butterfly diagrams of the axisymmetric azimuthal field $B_{\phi}^{m=0}$ for one example of each state. The magnetic field generated by the stationary dynamo is mostly axisymmetric and azimuthal (see the diagram in Fig.~\ref{fig:DipHem}(a)).
Unlike the other dynamo states, the strong dipolar branch is stable over large ranges of $N_o/\Omega_o$ and $Ro$. A stationary dynamo generating a field whose axisymmetric component has the opposite equatorial symmetry appears to be transient. Indeed, we eventually observe a transition to either the strong dipolar branch or the hemispherical branch,
where the magnetic energy concentrates in the southern or northern one depending on the initial condition. The butterfly diagram in Fig.~\ref{fig:DipHem}(b) shows that $B_{\phi}^{m=0}$ is not stationary and can locally change sign. Contrary to the stationary case, the energy associated with the non-axisymmetric magnetic field is of the same order of magnitude as that of the axisymmetric component. The hemispherical structure, which clearly emerges after a transient caused by the initial conditions, can be interpreted as the superposition of two dominant magnetic modes of dipolar and quadrupolar, with comparable energy in both modes~\cite{gallet2009,gallet2012}. In Fig.~\ref{fig:DipHem}, the bottom panel shows the time series of the parameter quantifying the equatorial symmetry breaking of the flow
\begin{equation}
    \Delta E_{\rm kin,H} \defeq \frac{E_{\rm kin,N}-E_{\rm kin,S}}{E_{\rm kin,N}+E_{\rm kin,S}}\,,
\end{equation}
which measures the difference between the kinetic energy in the northern ($E_{\rm kin,N}$) and the southern ($E_{\rm kin,S}$) hemispheres. We observe that it remains low in the stationary state, with $\Delta E_{\rm kin,H}\approx 0.15$, while the flow's equatorial symmetry is significantly broken in the hemispherical dynamo state, with $\Delta E_{\rm kin,H}\approx -0.5$. 

\begin{figure*}
    \centering
    \includegraphics[width=\textwidth]{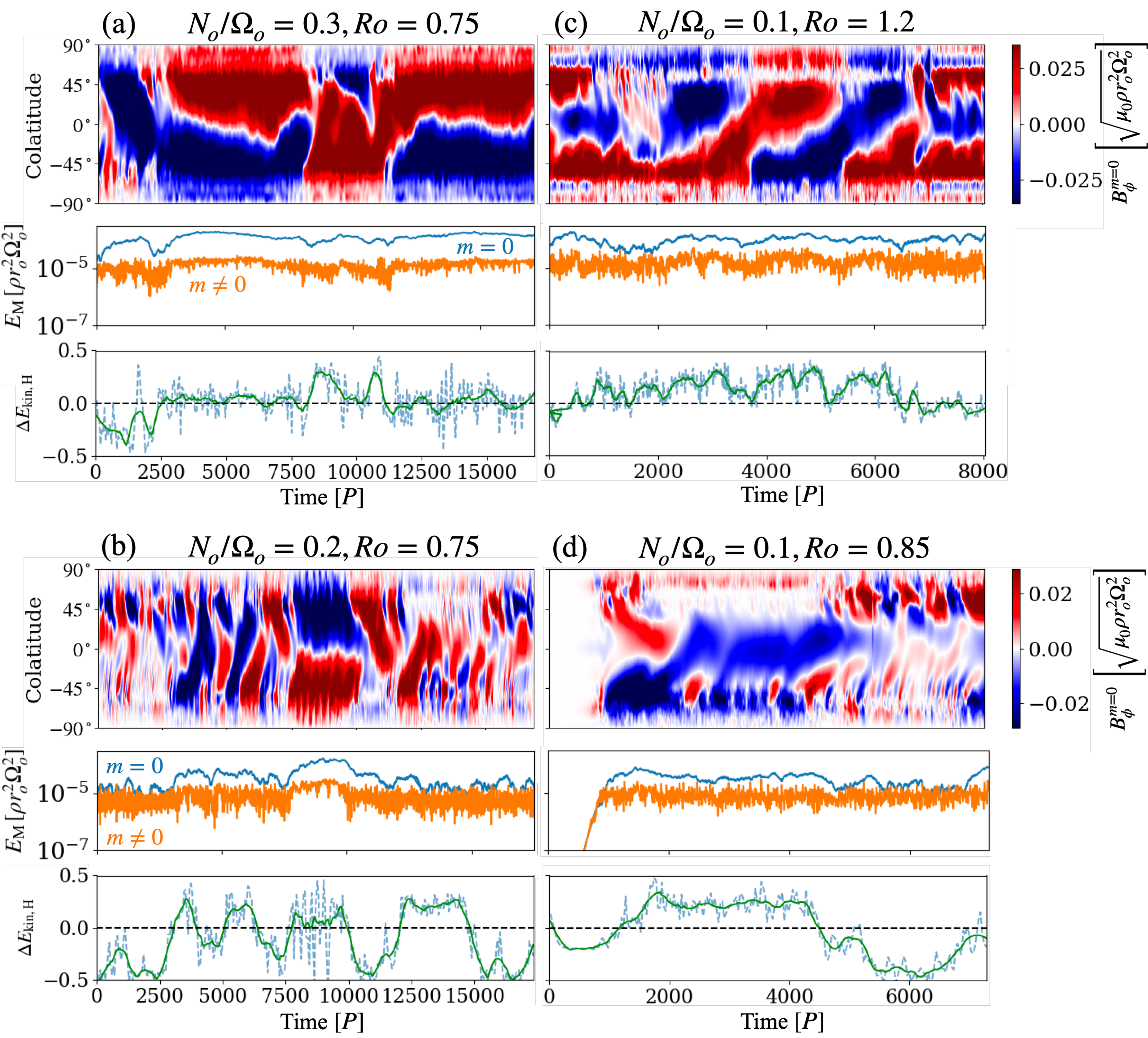}
    \caption{Same as Fig.~\ref{fig:DipHem} but for the parameters $N_o/\Omega_o=0.3,\,Ro=0.75$ (a), $N_o/\Omega_o=0.2,\,Ro=0.75$ (b), $N_o/\Omega_o=0.1,\,Ro=1.2$ (c), and $N_o/\Omega_o=0.1,\,Ro=0.85$ (d).}
    \label{fig:butterfly}
\end{figure*}

When $N_o/\Omega_o\in[0.175,0.3]$ or $Ro\geq0.85$, we observe random transitions between different dynamo regimes in the time series of several DNS. The resulting complex temporal dynamics is illustrated by the butterfly diagrams in Fig.~\ref{fig:butterfly}. For the cases at $N_o/\Omega_o=0.3, Ro=0.75$ and $N_o/\Omega_o=0.1, Ro=1.2$ (Figs.~\ref{fig:butterfly}(a, c)), the temporal dynamics is relatively slow. In Fig.~\ref{fig:butterfly}(a), the magnetic field remains dipolar and stationary during most of the simulation, but quick random reversals occur at the beginning of the DNS, at $t\approx 8000\,P$, and at $t\approx 10500\,P$. This is the first observation of magnetic field reversals produced by the Tayler–Spruit dynamo. In Fig.~\ref{fig:butterfly}(c), the magnetic field has an initial quadrupolar symmetry. We then observe a new dynamical regime, in which $B_{\phi}^{m=0}$ reverses in the southern hemisphere and migrates to the northern one. A more detailed description of this regime will be given in Sect.~\ref{ssec:physics}. Unlike in the hemispherical regime shown in Fig.~\ref{fig:DipHem}(b), the magnetic field remains predominantly axisymmetric, and eventually bifurcates onto the stationary regime with a quadrupolar magnetic field. In contrast to the transient case $Ro\leq0.85$, determining the stability of this quadrupolar state would require much longer integration times, and therefore remains uncertain.

When $N_o/\Omega_o=0.1, Ro=0.85$ and $N_o/\Omega_o=0.2, Ro=0.75$, the dynamics is more complex due to the transition to the hemispherical state. In Fig.~\ref{fig:butterfly}(d), the magnetic field location switches from the southern hemisphere to the northern one, but also becomes less axisymmetric after the change of hemisphere. Therefore, an axisymmetric hemispherical state exists, at least transiently, between $\approx1000$ and $\approx 4200\,P$. 

Moreover, we also observe for the first time a transition from the hemispherical to the strong dipolar branch, as illustrated in Fig.~\ref{fig:butterfly}(b). While initially in the northern hemisphere, the magnetic field starts reversing at $t\approx 6000\,P$, similarly to Fig.~\ref{fig:butterfly}(c). However, patches of $B_{\phi}^{m=0}$ travel toward the equator from either the northern or southern hemisphere, showing that the reversals can occur in both hemispheres at the same time. Finally, the magnetic field becomes transiently dipolar and axisymmetric, before reversing in the northern hemisphere and becoming hemispherical again. Note that the magnetic field location also changes from one hemisphere to the other at the end of the simulation ($t\approx 15000\,P$). This simulation shows that the magnetic field can switch to an axisymmetric magnetic field configuration despite starting from a non-axisymmetric geometry.

The butterfly diagrams in Fig.~\ref{fig:butterfly} are displayed with their associated time series of $\Delta E_{\rm kin,H}$. We first notice that the stationary states (axisymmetric dipolar or quadrupolar magnetic fields) are operating when $\Delta E_{\rm kin,H}\approx 0$, which is lower than in the stationary state at $N_o/\Omega_o=0.1,\,Ro=0.2$ (see Fig.~\ref{fig:DipHem}(b)). On the other hand, $|\Delta E_{\rm kin,H}|\gtrsim0.2$ when the magnetic field reverses or is localised in one hemisphere. Moreover, the switch of the magnetic field location corresponds to the sign changes of $\Delta E_{\rm kin,H}$: the magnetic field is located in the northern (southern) hemisphere when $\Delta E_{\rm kin,H}<0$ ($\Delta E_{\rm kin,H}>0$). Therefore, the temporal dynamics of the magnetic field is strongly related to the  equatorial symmetry breaking of the flow that occurs spontaneously in our DNS. 

Finally, the dynamo is more difficult to sustain for $Ro=0.75,\,N_o/\Omega_o\geq0.5$. To capture it, we therefore increased $Pm$ to $2$, and then $4$. In this regime, the Tayler-Spruit dynamo shows an intermittent behaviour, in which the Tayler modes are triggered and damped quasi-periodically. This state could be explained by the proximity to the instability threshold~\citep{barrere2024a}.

\section{Interpretation of the temporal dynamics in one DNS}\label{sec:cycle}

\begin{figure*}
    \centering
    \includegraphics[width=0.75\textwidth]{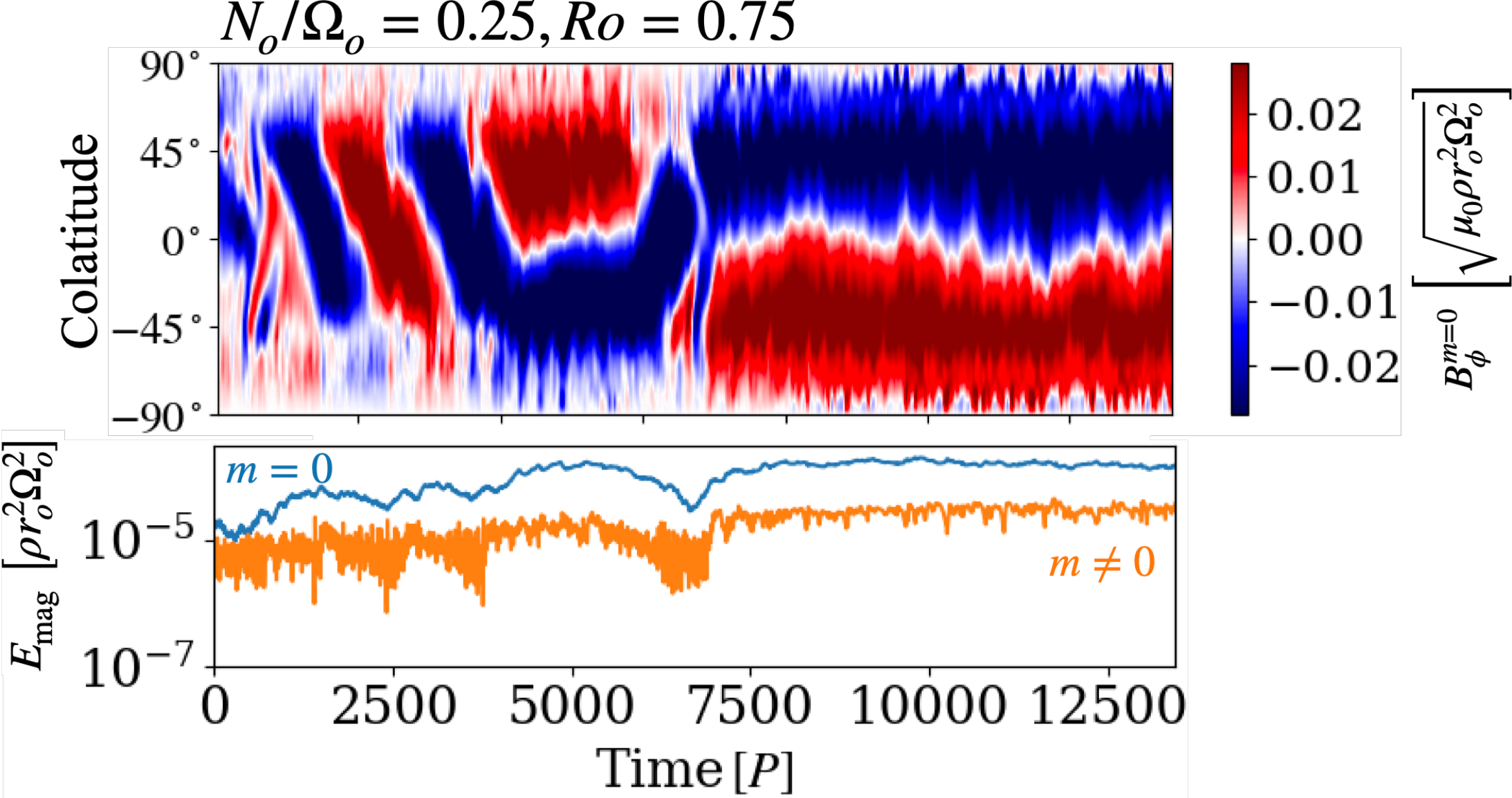}
    \caption{Butterfly diagram of the axisymmetric azimuthal magnetic field averaged over the radius interval $[0.42,0.5]\,r_o$ (top), and time series of the axisymmetric and non-axisymmetric magnetic energies (bottom) for the DNS at $N_o/\Omega_o=0.25,\,Ro=0.75$.}
    \label{fig:butt_NO025}
\end{figure*}

While we observed various complex temporal dynamics in the previous section, we propose here an interpretation of the Tayler-Spruit dynamo behaviour at $Pm=1\,,Ro=0.75\,,N_o/\Omega_o=0.25$. The focus on this DNS is motivated by the relatively simple dynamics: the magnetic field is axisymmetric, and we observe a clear transition from a reversing magnetic field to a stationary dipolar geometry, as seen in Fig.~\ref{fig:butt_NO025}. We will first describe in detail the dynamics of the reversals and then identify the bifurcation through which this state emerges.

\begin{figure*}
    \centering
    \includegraphics[width=\textwidth]{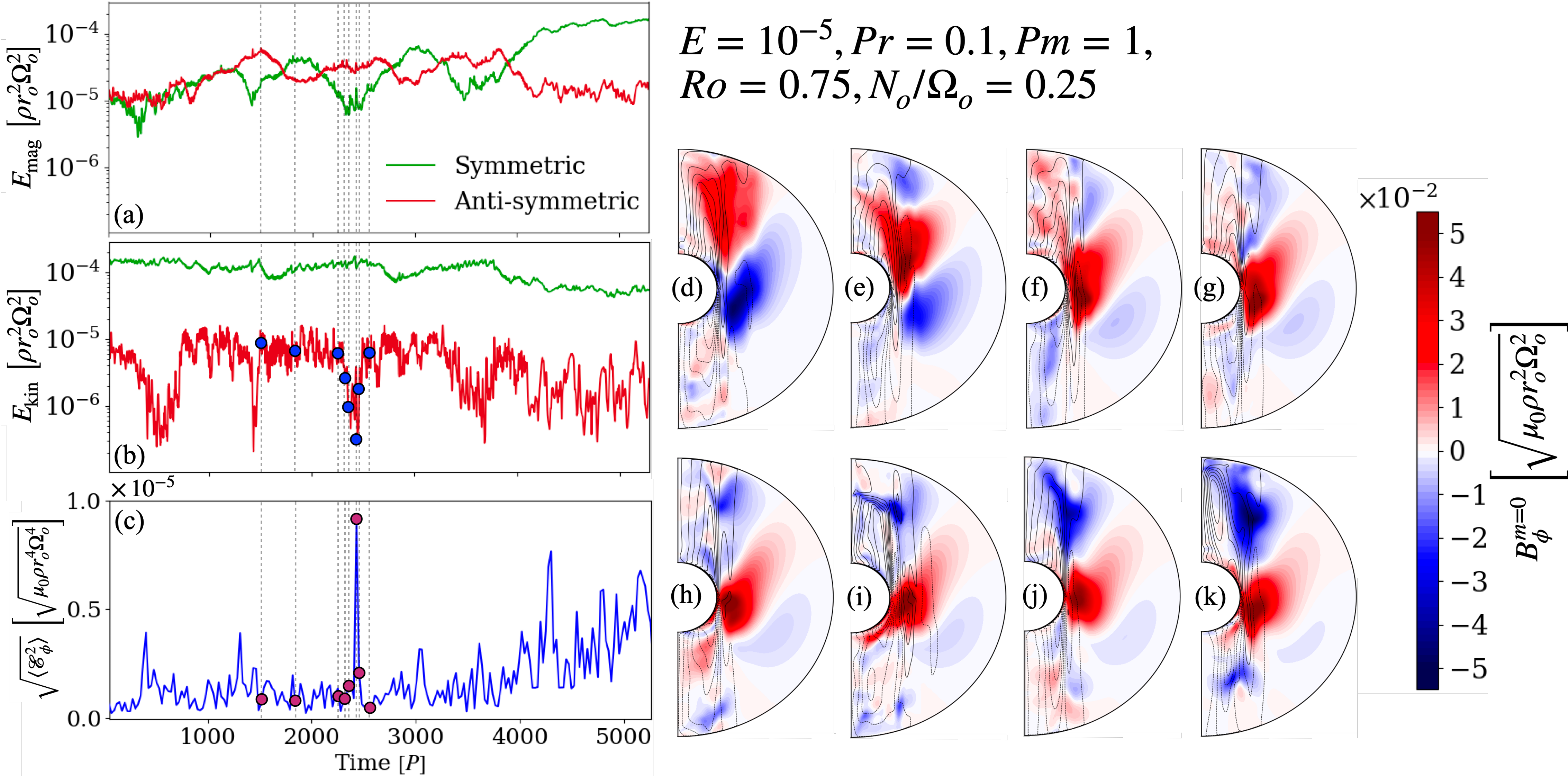}
    \caption{Left: Time series of the equatorially symmetric and antisymmetric components of (a) the magnetic and (b) kinetic energies, and of (c) the root-mean-square azimuthal electromotive force. Right: Meridional slices of the axisymmetric azimuthal magnetic field and the isocontours of the meridional circulation (black solid/dashed lines) at times: (d) $t=1591\,P$; (e) $t=1919\,P$; (f) $t=2303\,P$; (g) $t=2355\,P$; (h) $t=2405\,P$; (i) $t=2430\,P$; (j) $t=2481\,P$; (k) $t=2609\,P$. They are associated with the blue and purple circles, and vertical dashed grey lines on the time series.}
    \label{fig:ts_dynamics}
\end{figure*}

\subsection{Description of the dynamics}\label{ssec:physics}

\begin{figure*}
    \centering
    \includegraphics[width=0.5\textwidth]{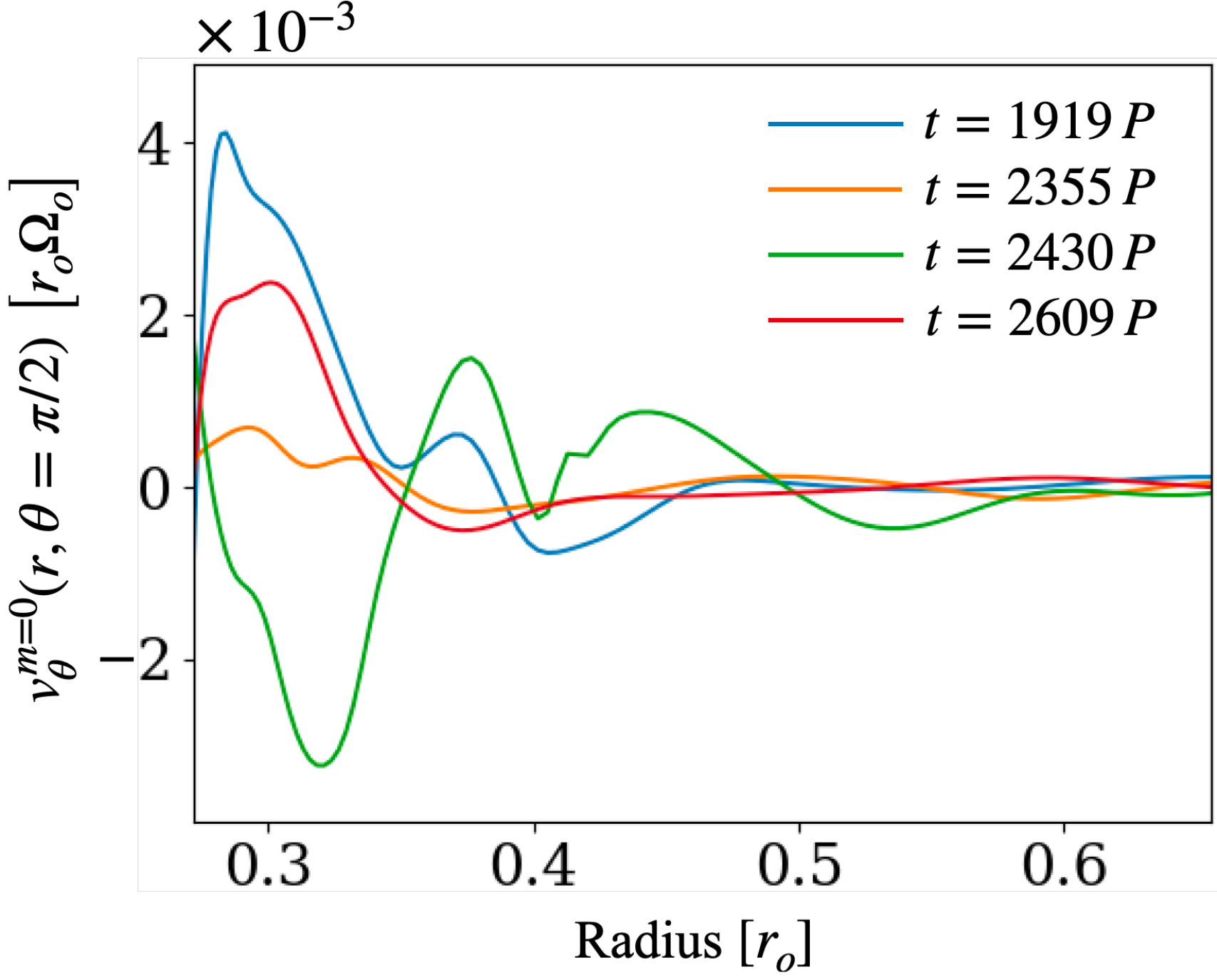}
    \caption{Radial profiles of the axisymmetric latitudinal velocity at the equator for the DNS with parameters $Pm=1,\,Ro=0.75,\,N_o/\Omega_o=0.25$.}
    \label{fig:vs_dynamics}
\end{figure*}

Figures~\ref{fig:ts_dynamics}(a, b) show how the magnetic and kinetic energies associated with both equatorial symmetries evolve during the reversing state, which occurs between $t=1000\,P$ and $t=4000\,P$. We clearly observe oscillations of both symmetries of the magnetic energy, which evolve in antiphase. This periodic transfer of energy between both parities was already observed in numerical simulations of the convective dynamo~\cite{raynaud2016} and can be described as the Type 1 modulation defined by~\citet{knobloch1998}. The minima of the symmetric magnetic energy (and so the maxima of the antisymmetric component) coincide with the sharp drops of the antisymmetric kinetic energy by a factor $\approx 50$ and slight increases of the symmetric component. 
These time series can be explained by looking at the meridional slices of the axisymmetric azimuthal magnetic field, $B^{m=0}_{\phi}$, in Figs.~\ref{fig:ts_dynamics}(d-k). Here, we focus on one period of the magnetic energy signal between $t=1591\,P$ and $t=2609\,P$, and each snapshot corresponds to a vertical dashed grey line and a blue circle on the curve of the antisymmetric kinetic energy. In Fig.~\ref{fig:ts_dynamics}(d), $B^{m=0}_{\phi}$ is concentrated in the northern hemisphere and composed of two large-scale patches of opposite polarities. The concentration of negative $B^{m=0}_{\phi}$ (blue in the snapshots) is located at the equator, hence the maxima in antisymmetric magnetic energy. Both patches then migrate towards the equator, until the negative concentration passes the equator (hence the maxima in symmetric magnetic energy, Fig.~\ref{fig:ts_dynamics}e), decays and eventually disappears as it is no longer fed by the Tayler-Spruit dynamo (Fig.~\ref{fig:ts_dynamics}f). The migration might be partially due to advection by the meridional circulation in the northern hemisphere (black lines in the snapshots), whose cell crosses the equator. This equatorial asymmetry of the meridional circulation is clear in Fig.~\ref{fig:vs_dynamics}, which shows the radial profile of the axisymmetric component of the latitudinal velocity at the equator at different times during one magnetic field reversal. The blue curve at $t=1919\,P$ shows a peak in velocity near the inner sphere due to the northern meridional circulation cell. The asymmetry is certainly produced by the nonlinear interaction between the Tayler instability and the flow, as it also appears on the hemispherical dynamo branch, but not in the hydrodynamical Taylor-Couette flow simulations with the same parameters. 

Between $t=2355\,P$ and $t=2405\,P$ (Figs.~\ref{fig:ts_dynamics}g, h), the meridional circulation cell no longer crosses the equator, which translates into a flat radial profile of $v_\theta^{m=0}$ at the equator at $t=2355\,P$ (orange line in Fig.~\ref{fig:vs_dynamics}).  As a consequence, the flow symmetrises, as shown by the sudden decrease of the antisymmetric kinetic energy. The concentration of positive  $B^{m=0}_{\phi}$ (red patch in the snapshots) therefore remains at the equator. At $t=2430\,P$ (Fig.~\ref{fig:ts_dynamics}i), a negative concentration of $B^{m=0}_{\phi}$ forms in the northern hemisphere and the magnetic field tends towards the same geometry as in Fig.~\ref{fig:ts_dynamics}(a), but with the opposite polarity (Figs.~\ref{fig:ts_dynamics}j, k). The same process repeats to produce a second reversal and form a cycle at $t=3800\,P$. After this time, the concentration of toroidal field has migrated through the equator and is not dissipated. The magnetic field therefore has a dipolar symmetry and becomes stationary. 

\begin{figure*}
    \centering
    \includegraphics[width=0.8\textwidth]{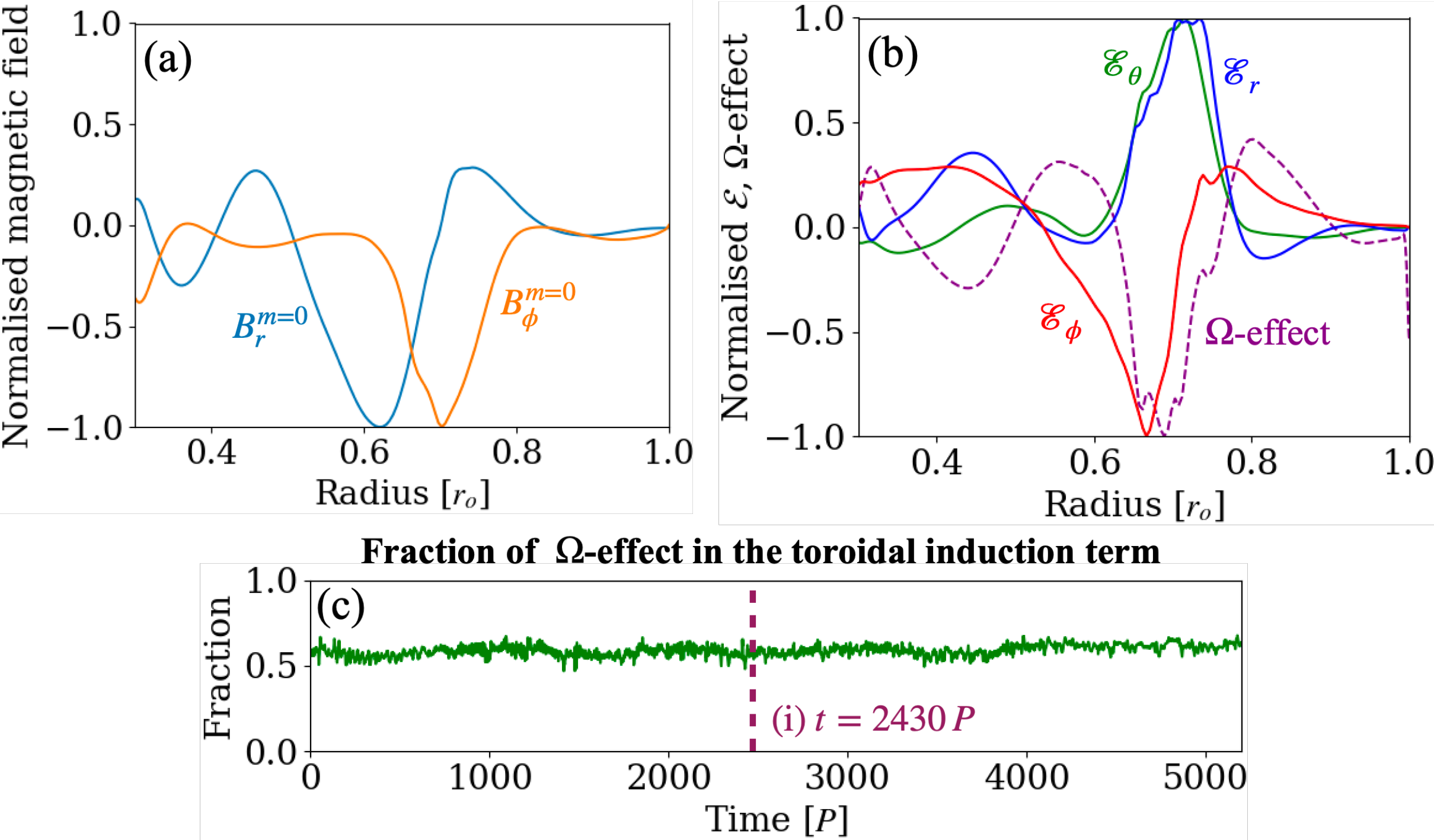}
    \caption{Radial profiles averaged between the colatitudes $[11^{\circ},25^{\circ}]$ at $t=2430\,P$ of (a) the axisymmetric radial and azimuthal magnetic fields, (b) the components of the electromotive force, and the $\Omega$-effect. The quantities are normalised by their maximum values. (c) Time series of the fraction of $\Omega$-effect (azimuthal component of $\nabla\times(v_{\phi}\mathbf{e}_{\phi}\times\mathbf{B})$~\citep{schrinner2011}) in the toroidal induction term (azimuthal component of $\nabla\times(\mathbf{v}\times\mathbf{B})$). The purple horizontal bar indicates the time of the electromotive burst visible in Fig.~\ref{fig:ts_dynamics}.}
    \label{fig:induction}
\end{figure*}

Looking at the time series of the root-mean-squared and volume-averaged azimuthal electromotive force (EMF) in Fig.~\ref{fig:ts_dynamics}(c), we observe a burst at $t=2430\,P$, where the azimuthal EMF, $\mathcal{E}_{\phi}$, increases suddenly by a factor of $\approx 8$. It coincides with the drop in antisymmetric kinetic energy and the minimum in symmetric magnetic energy. This term, which is produced by the non-axisymmetric dynamics of the Tayler instability, is usually responsible for the formation of the axisymmetric radial magnetic field. The axisymmetric azimuthal component $B^{m=0}_{\phi}$ is eventually produced by the $\Omega$-effect, namely the shearing of the radial magnetic field. We see in Figs.~\ref{fig:induction}(a, b) that the axisymmetric field location in the northern hemisphere is correlated to the peaks of $\mathcal{E}_{\phi}$ and of the $\Omega$-effect. This suggests that the emerging concentration of $B_{\phi}$ in the northern hemisphere is formed via this $\mathcal{E}_{\phi}$-$\Omega$ process. However the radial and latitudinal components of the EMF, $\mathcal{E}_{r}$ and $\mathcal{E}_{\theta}$ also participate in the induction of $B_{\phi}^{m=0}$. Fig.~\ref{fig:induction}(b) also shows these quantities are spatially anti-correlated with $B_{\phi}^{m=0}$. Both processes are equally important because the time series in Fig.~\ref{fig:induction}(c) indicates that the $\Omega$-effect represents $\approx\SI{55}{\%}$ of the total induction at the time of the peak in $\mathcal{E}_{\phi}$.Radial profiles of the axisymmetric latitudinal velocity at the equator for the DNS with parameters $Pm=1,\,Ro=0.75,\,N_o/\Omega_o=0.25$.

Figure~\ref{fig:Bp_dynamics} displays the trajectory of the system in the reduced phase space consisting of the plane $B_{\phi}^{\rm Equator}$-$B_{\phi}^{\rm North}$. These two quantities are defined in the meridional slice on the right of Fig.~\ref{fig:Bp_dynamics} as the axisymmetric azimuthal magnetic field averaged over the hatched regions ($[0.345 r_0,\,0.395 r_0]$,\,[85°,\,95°]) for $B_{\phi}^{\rm Equator}$, and ($[0.67 r_0,\,0.72 r_0]$,\,[23°,\,33°]) for $B_{\phi}^{\rm North}$. This plot gives us another representation of the cycle, which forms a closed orbit. On this diagram, we can follow the trajectory on the half cycle illustrated by Fig.~\ref{fig:ts_dynamics}(d-k), starting from the blue circle (d) located at $(-0.05,0.03)$. The system goes towards the line $B^{\rm North}_{\phi}=0$ as the azimuthal magnetic field migrates towards the equator. The dynamics becomes very slow around $(0.05,0)$, until the burst of $\mathcal{E}_{\phi}$ where the system rapidly reaches $(0.05,-0.03)$ as $B^{\rm North}_{\phi}$ is regenerated with the opposite sign. The speed of the trajectory in this cycle recalls, for instance, the hydrodynamics of states involved in the transition between laminar and turbulent flows in the asymptotic suction boundary layer (ASBL) studied by~\citet{kreilos2013}. They find that the trajectory speeds up during bursts of the cross-flow energy (used as an indicator for the presence of turbulence) while completing one half of the orbit, and significantly slows down while approaching the foot of the next burst. To determine the nature of the points where the speed trajectory is minimum in the phase space (e.g. presence of a weakly unstable state, or a near collision of a stable and an unstable fixed point), we need to understand how the cycle emerges.

\subsection{Role of the flow equatorial symmetry}\label{ssec:dynamics}

\begin{figure*}
    \centering
    \includegraphics[width=\textwidth]{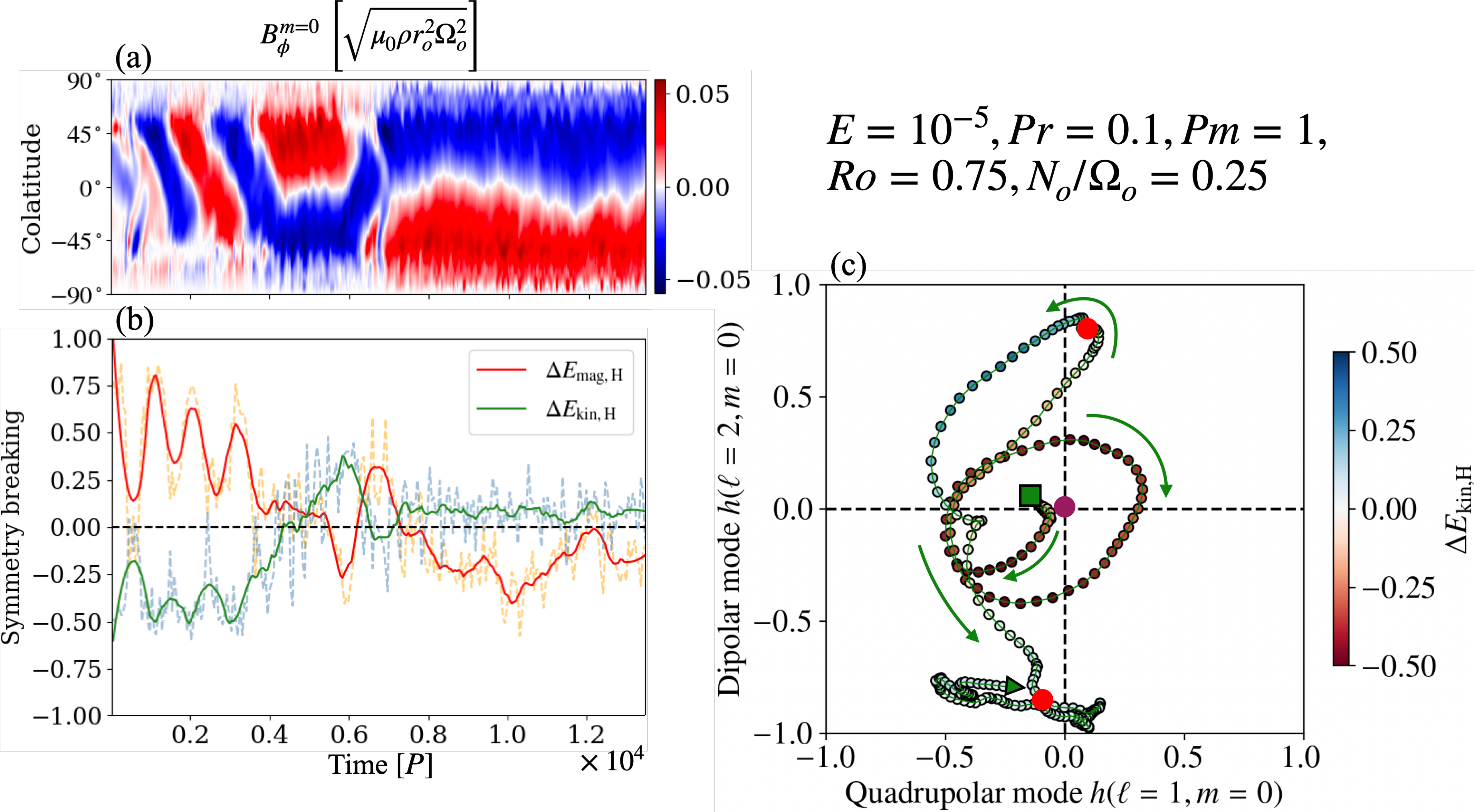}
    \caption{(a) Butterfly diagram of the axisymmetric azimuthal magnetic field averaged over the radial interval $[0.42,0.5]\,r_o$. (b) Time series of the equatorial symmetry breaking of the magnetic (orange dashed line; red solid line for the smoothed signal) and kinetic (blue dashed line; green solid line for the smoothed signal) energies. (c) Smoothed trajectory of the system in the dipolar--quadrupolar plane. The colormap of the scatter plot indicates the value of the equatorial symmetry breaking of the kinetic energy. The green arrows indicate the direction of the trajectory, and the green square and triangle mark the start and the end of the simulation. The red and violet circles represent the fixed points associated with the strong stationary state and the non-dynamo state, respectively. The strengths of the dipolar and quadrupolar modes are both normalised by dividing by the absolute maximum strength of the dipolar mode.}
    \label{fig:fiducial}
\end{figure*}

While we focused on the cycle we captured in the simulation at $Pm=1\,,Ro=0.75\,,N_o/\Omega_o=0.25$ in Sect.~\ref{ssec:physics}, we will now investigate the bifurcation to the stationary dipolar state. The butterfly diagram of $B^{m=0}_{\phi}$ displayed in Fig.~\ref{fig:butt_NO025} and Fig.~\ref{fig:fiducial}(a) shows that the dynamo generates magnetic field reversals until $t=4000\,P$, and then a stationary dipolar field. Finally, $B^{m=0}_{\phi}$ polarity changes with a reversal between $t=5800\,P$ and $t=7000\,P$, before remaining stationary. 
The observed correlation between $\Delta E_{\rm kin,H}$ and the spatio-temporal dynamics of the dynamo is reminiscent of a framework initially introduced by Pétrélis \& Fauve~\cite{petrelis2008chaotic} in the context of the VKS dynamo experiment to explain the emergence of randomly reversing~\cite{petrelis2008chaotic,petrelis2009,gallet2012reversals} and/or hemispherical magnetic fields~\cite{gallet2009,gallet2012}. Indeed, one generically expects nonzero $\Delta E_{\rm kin,H}$ to couple
two large-scale dynamo modes with opposite equatorial symmetries: a dipolar and a quadrupolar mode.
During the reversing phase, we see in Fig.~\ref{fig:fiducial}(b) that $\Delta E_{\rm kin,H}\approx -0.4$, which is consistent with the reversing states observed in Sect.~\ref{sec:param}. We also note oscillations where $\Delta E_{\rm kin,H}$ tends towards $\Delta E_{\rm kin,H}=0$, which is correlated with the symmetrisation of the flow observed just before the EMF bursts during the magnetic field reversal. After $t=4000\,P$, the flow becomes equatorially symmetric, which coincides with the first transition towards the strong stationary state. The flow equatorial symmetry is again broken as $\Delta E_{\rm kin,H}$ keeps increasing to reach $\approx 0.3$ at $6000\,P$ (a value comparable to that observed during hemispherical reversals in the interval $t\in[1000\,P,4000\,P]$ but in the southern hemisphere), leading to another reversal. Finally, the flow symmetrises one last time as $\Delta E_{\rm kin,H}$ decreases down to $\approx0.1$ and remains quasi-constant. This matches with the second transition to the dipolar stationary state, which remains from $t\approx 6500\,P$ until the end of the simulation. The correlation between the flow symmetry breaking and the different dynamical phases is therefore confirmed.

\begin{figure*}
    \centering
    \includegraphics[width=\textwidth]{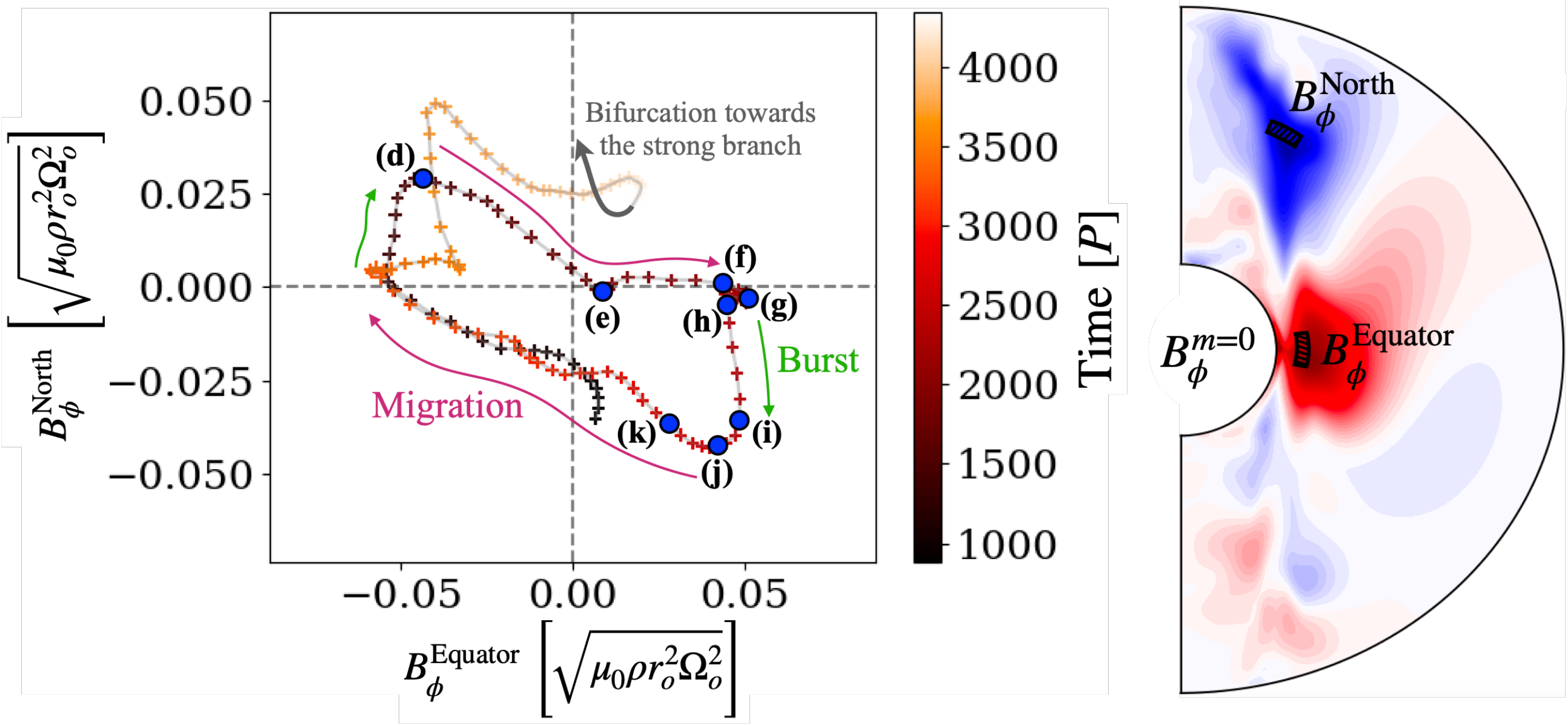}
    \caption{Phase diagram in the plane $B_{\phi}^{\rm Equator}$-$B_{\phi}^{\rm North}$, which are defined in the meridional slice on the right as the axisymmetric azimuthal magnetic field averaged over the hatched regions ($[0.345 r_0,\,0.395 r_0]$,\,[85°,\,95°]) for $B_{\phi}^{\rm Equator}$, and ($[0.67 r_0,\,0.72 r_0]$,\,[23°,\,33°]) for $B_{\phi}^{\rm North}$. The colour of the crosses represents the time between $\approx900\,P$ and $\approx4250\,P$. The blue circles represent the same times as in Fig.~\ref{fig:ts_dynamics}(b). The arrows indicate the direction of the trajectory.}
    \label{fig:Bp_dynamics}
\end{figure*}

Since the magnetic field and the flow impact each other when the former is strong enough, as in the Tayler-Spruit dynamo, it is also interesting to look at the symmetry breaking of the magnetic field, which is equivalent to the hemisphericity of the dynamo. For that purpose, we define a parameter $\Delta E_{\rm mag,H}$ similar to $\Delta E_{\rm kin,H}$:
\begin{equation}
    \Delta E_{\rm mag,H} \defeq \frac{E_{\rm mag,N}-E_{\rm mag,S}}{E_{\rm mag,N}+E_{\rm mag,S}}\,.
\end{equation}
The evolution of this parameter is shown as the red curve in Fig.~\ref{fig:fiducial}(b).
However, we observe that $\Delta E_{\rm kin,H}$ and $\Delta E_{\rm mag,H}$ are in anti-phase, namely the magnetic field is located in the hemisphere with the lower kinetic energy. A similar observation was made in the VKS dynamo experiment~\cite{gallet2012}. In the VKS context, the symmetry breaking is imposed externally through the prescribed rotation speeds of the impellers. In the present simulations, by contrast, the symmetry breaking arises spontaneously, possibly enhanced by the hemispherical field that further suppresses the flow in the magnetically active hemisphere.

We now extract the dipolar/quadrupolar modes by expanding the toroidal magnetic potential $h$ into spherical harmonics. In most simulations, the dominant equatorially symmetric and antisymmetric components are the $\ell=2$ and $\ell=1$ modes, respectively. Therefore, we define the dipolar and the quadrupolar modes as $h(\ell=2,m=0)$ and $h(\ell=1,m=0)$, respectively. While the term `dipole' or `dipolar' can be associated to the spherical harmonic degree $\ell=1$, we recall that this definition is relevant for the poloidal field and not the toroidal field, which has an opposite equatorial symmetry (quadrupolar). The trajectory of the magnetic field in the 2D dipolar-quadrupolar phase space is shown in Fig.~\ref{fig:fiducial}(c) with a colour map illustrating the value of $\Delta E_{\rm kin,H}$. While the magnetic field reverses between $t\approx1000-4000\,P$, it moves clockwise along a limit cycle during which $\Delta E_{\rm kin,H}\approx 0.3-0.5$. When $\Delta E_{\rm kin,H}$ tends toward $0$, the magnetic field then increases toward a first fixed point (red circle at $(0.07,0.84)$) where the magnetic field dynamics slows down. As $\Delta E_{\rm kin,H}$ increases, the magnetic field reverses and so moves back toward the cycle. As soon as it reaches the cycle, $\Delta E_{\rm kin,H}$ decreases, and the magnetic field eventually connects to the $B \to -B$ symmetric fixed point (red circle at $(-0.07,-0.84))$. One may have expected this part of the trajectory to be in the right-hand region of the phase space $h(\ell=2,m=0)>0$ and to travel clockwise like on the limit cycle. It actually goes in the opposite direction because the reversal happens in the opposite hemisphere (see the time series in Fig.~\ref{fig:fiducial}(b), consistent with the change in colorscale in Fig.~\ref{fig:fiducial}(c)). Finally, the magnetic field fluctuates around the fixed point and $\Delta E_{\rm kin,H}$ remains quasi-constant. 

\begin{figure*}
    \centering
    \includegraphics[width=\textwidth]{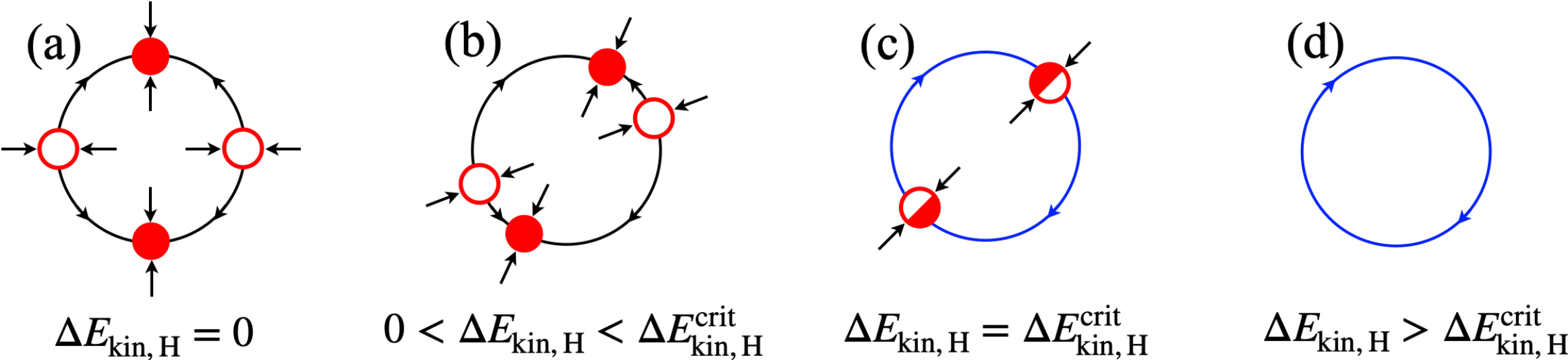}
    \caption{Schematic 2D representation of the saddle-node on invariant cycle bifurcation in the state space. (a) The fixed points associated with the stable dipolar state (red filled circles) and the unstable quadrupolar state (red empty circles) coexist and are heteroclinically connected. (b) As the flow equatorial symmetry breaking increases, the fixed points with opposite symmetries come closer to each other. (c) The fixed points collide. (d) The fixed points are gone, and the system follows a limit cycle.} \label{fig:sniper}
\end{figure*}

The trajectory is consistent with the emergence of a limit cycle due to the collision of two pairs of fixed points that are heteroclinically connected~\cite{petrelis2008chaotic,petrelis2009}. The structure of this bifurcation is illustrated in Fig.~\ref{fig:sniper}. When $\Delta E_{\rm kin,H}=0$, the dipolar--quadrupolar space contains two pairs of stable fixed points (red filled circles)/saddle points (red empty circles) with opposite signs, which are connected heteroclinically. In our case, the stable mode is the dipolar mode because the stable fixed points are located near the axis $h(l=1,m=0)=0$. As the equatorial symmetry of the flow is broken ($\Delta E_{\rm kin,H}\neq 0$), the dipolar and quadrupolar modes of the same sign draw closer and closer. Once they collide, the magnetic field follows the former heteroclinic connections in a limit cycle. In the framework of this bifurcation, the slow dynamics around the points at $B^{\rm North}_{\phi}=0$ in Fig.~\ref{fig:Bp_dynamics} --- which is less visible in the diagram in Fig.~\ref{fig:fiducial}(c) --- can be interpreted as the `ghosts' resulting from the collision of the fixed points with the saddle points on the cycle. This is a typical characteristic of a saddle-node on invariant cycle (SNIC) bifurcation ~\citep{tuckerman1988,abshagen2005,kreilos2013}. We stress that this bifurcation only explains the transition between the reversing and the stationary dipolar states, and does not include the presence of the hydrodynamic stable fixed point at $(0,0)$ due to the subcritical nature of the Tayler-Spruit dynamo.

\begin{figure*}
    \centering
    \includegraphics[width=\textwidth]{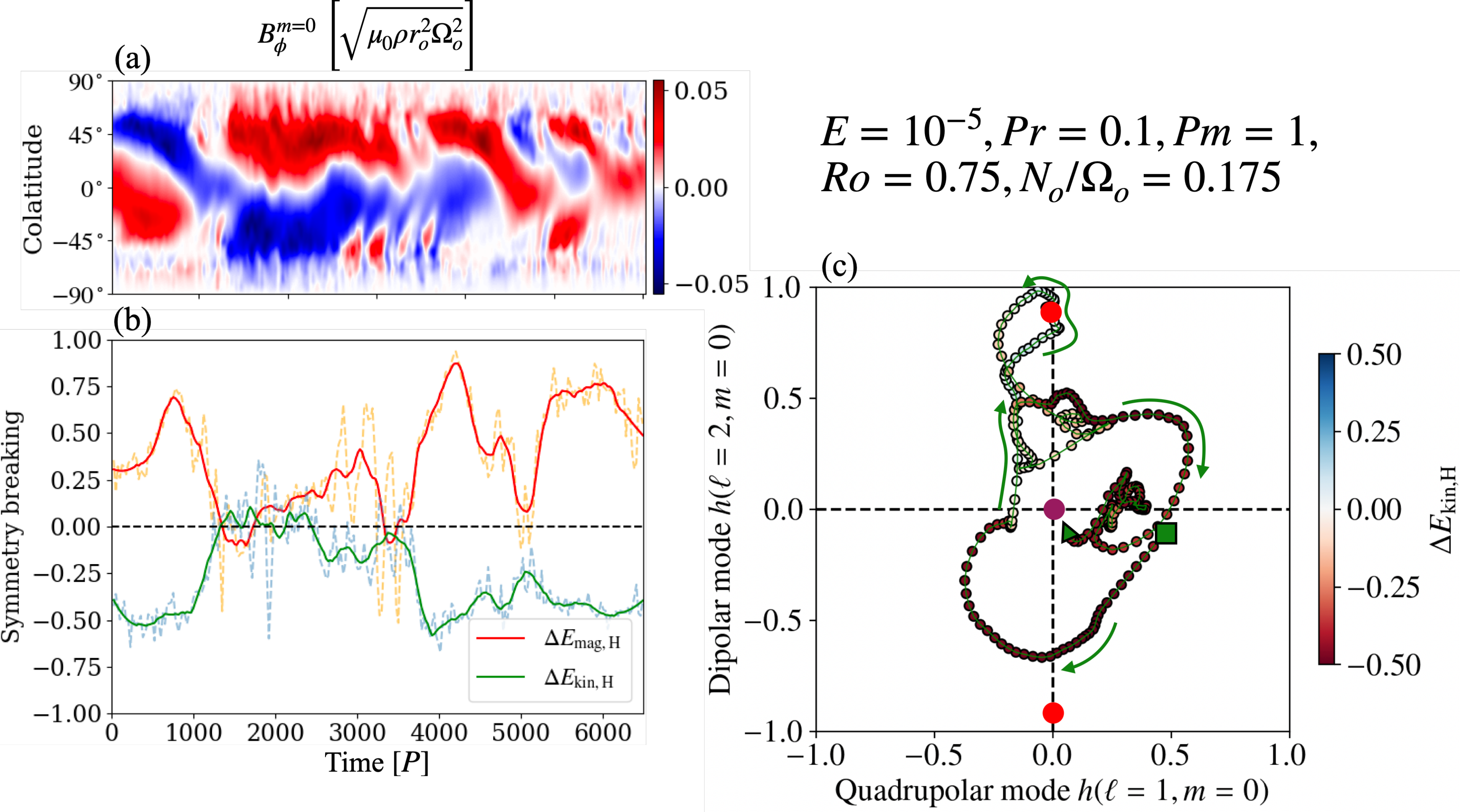}
    \caption{Same as Fig.~\ref{fig:fiducial} but for $N_o/\Omega_o=0.175$.}
    \label{fig:fiducialBIS}
\end{figure*}

Another simulation ($Pm=1$, $Ro=0.75$, $N_o/\Omega_o=0.175$) exhibits two reversals (at $\approx1000\,P$ and $\approx4500\,P$) that follow a limit-cycle, before settling in a hemispherical dynamo state as seen in the butterfly diagram of Fig~\ref{fig:fiducialBIS}. The observed behaviour is broadly compatible with the two-magnetic-mode description of the system we have presented. Indeed, a significant symmetry breaking of the flow ($\Delta E_{\rm kin,H}\approx -0.5$) coincides with a reversing state, while the stationary state coincides with $\Delta E_{\rm kin,H}\approx 0$ (see Fig.~\ref{fig:fiducialBIS}(b-c)). Interestingly, the final hemispherical state also coincides with a symmetry breaking of the flow, like the reversing state. \citet{gallet2009} already predicted that the coupling of two dynamo modes with opposite parities can either produce reversals or magnetic field localisation depending on the turbulent fluctuations that break the flow symmetry. Beyond such low-dimensional dynamics, however, this simulation, and those displayed in Sect.~\ref{sec:param}, also display more complex magnetic fluctuations, which highlights the likely influence of smaller-scale magnetic modes expected in the $Pm=\mathcal{O}(1)$ regime.

\begin{figure*}[t!]
    \centering
    \includegraphics[width=0.95\textwidth]{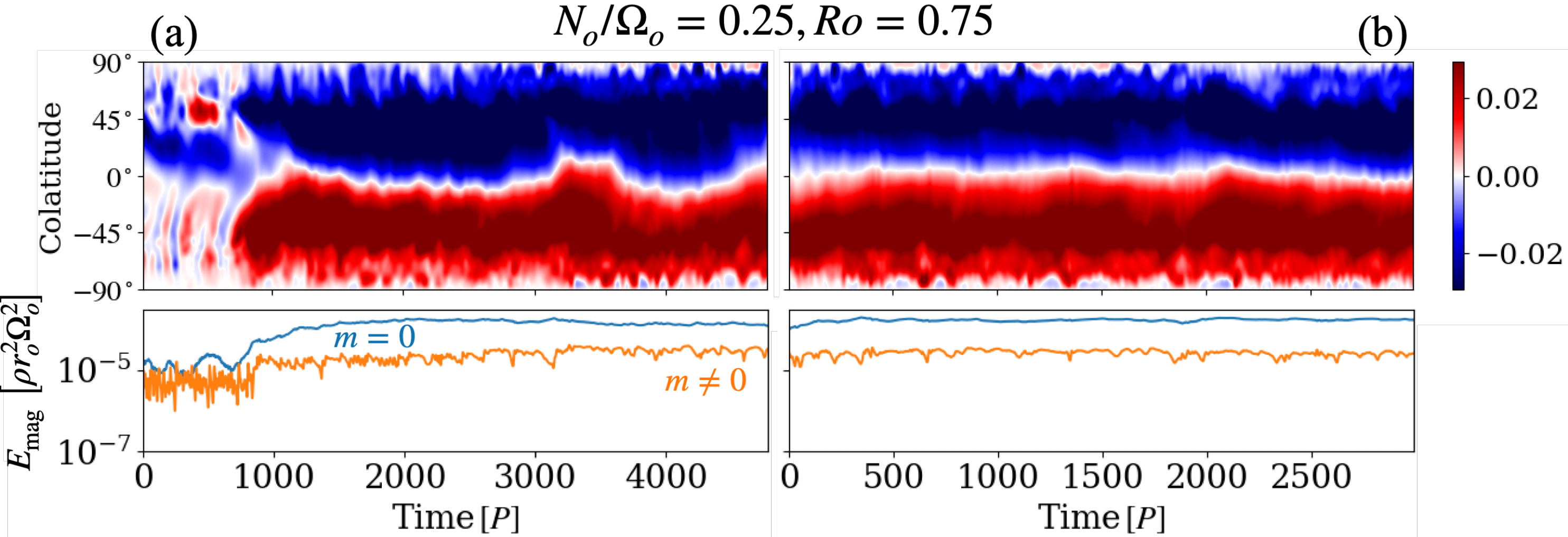}
    \caption{Butterfly diagram of the axisymmetric azimuthal magnetic field averaged over the radius interval $[0.42,0.5]\,r_o$ (top). Time series of the axisymmetric and non-axisymmetric magnetic energies (bottom) for two DNS at $N_o/\Omega_o=0.25,\,Ro=0.75$, but with different initial conditions: a hemispherical state at $N_o/\Omega_o=0.2,\,Ro=0.75$ (a) and a dipolar state at $N_o/\Omega_o=0.1,\,Ro=0.75$ (b).} \label{fig:butt_NO025BIS}
\end{figure*}

Finally, we note that the trajectory in phase space is also sensitive to the initial conditions, especially for the same parameters as those of the DNS we have investigated in this section ($Pm=1,\,Ro=0.75,\,N_o/\Omega_o=0.25$). For instance, starting from a hemispherical state obtained for a DNS with $N_o/\Omega_o=0.1$ (see Fig.~\ref{fig:butt_NO025BIS}(a)), the symmetry breaking quickly reaches and stays at $\Delta E_{\rm kin,H}=0$. The system therefore bifurcates from the hemispherical to the strong branch without displaying any transient reversals. This example indicates the difficulty in capturing the reversing state. Unsurprisingly, when we start from the strong stationary state at $N_o/\Omega_o=0.1$, and so using initial conditions close to a stable fixed point at $\Delta E_{\rm kin,H}=0$, the Tayler-Spruit dynamo remains in the same stationary state (see Fig.~\ref{fig:butt_NO025BIS}(b)). We also observe this sensitivity to the initial conditions for $Pm=1,\,Ro=0.75,\,N_o/\Omega_o=0.3$.

\section{Conclusions}\label{sec:conclu}
The numerical investigation presented in this paper shows for the first time that the Tayler-Spruit dynamo can generate a wide variety of dynamical regimes, including stationary magnetic fields with different equatorial symmetries, hemispherical location, or magnetic field reversals. The temporal dynamics can also be very complex, with several transitions between the different regimes. They are correlated with the evolution of the equatorial symmetry of the flow, which is broken spontaneously. This result deepens our understanding of the physics behind the Tayler-Spruit dynamo. Additionally,  longer DNS will be needed to further characterise the rich dynamical landscape of this dynamo.

We also propose an interpretation to explain the `simple' temporal dynamics in one fiducial DNS, in which the magnetic field only transitions between a reversing state and a stationary state with a dipolar symmetry. From a physical point of view, the reversal occurs in two phases: (i) the slow migration of same-polarity magnetic field towards the equator (likely partially attributable to advection by the meridional circulation cell), and (ii) the burst of electromotive force associated with the fast formation of large-scale magnetic fields in one hemisphere with the opposite polarity. This translates into a limit cycle in phase space on which the evolution is characterised by a slow and a fast timescale. From the point of view of dynamical systems, the cycle can be interpreted as emerging from the collision of two symmetry-related pairs of stable and unstable fixed points, when the equatorial symmetry of the flow is broken. Namely, the symmetry breaking of the flow couples two large-scale magnetic modes with opposite equatorial symmetries. Similar global bifurcations have already been identified in different types of hydrodynamical~\citep{kreilos2013} or magnetic flows~\citep{petrelis2009}.

It is interesting to compare this interpretation with the dynamics generated by the dynamo in VKS. The first striking similarity is the emergence of a reversing magnetic field when the flow equatorial symmetry is broken. In contrast, a key difference is that the symmetry breaking spontaneously emerges in our system, whereas it is a control parameter of the VKS experiment. The consequence is that the cycle period cannot be made arbitrarily large in our setup. Instead, it is set by the typical magnitude of the spontaneously emerging symmetry breaking of the flow. This comparison between two very different setups supports the idea that symmetry breaking of the flow is a key ingredient associated with magnetic field reversals.
\citet{petrelis2009} derive a low-dimensional model coupling two large-scale dynamo modes with opposite symmetries to explain the different dynamical regimes generated by the dynamo in VKS. Restricting the dynamics to two interacting dynamo modes is particularly justified for VKS because the system operates close to the threshold of a supercritical dynamo bifurcation. Moreover, the VKS dynamo operates in a low-magnetic-Prandtl-number regime ($Pm\approx 10^{-5}$)~\citep{monchaux2007} where small-scale magnetic modes are slaved to the dynamics of the dominant dipolar and quadrupolar modes. In our system, by contrast, the dynamo is reached subcritically and $Pm\approx 1$, which is several orders of magnitude larger than in VKS. In the low-$Pm$ VKS dynamo, the symmetry-breaking velocity perturbations pre-exist the emergence of the dynamo magnetic field, and the velocity symmetry-breaking thus 'causes' the magnetic reversals. By contrast, in the present $Pm\approx 1$ system, we expect a strong two-way coupling between the magnetic field and the symmetry-breaking velocity perturbations, making the causal direction between these two phenomena more ambiguous. Beyond such a two-magnetic-mode interpretation, the larger $Pm$ of the present system also suggests that more than two dynamo modes may be interacting, thus explaining the greater variety of spatio-temporal dynamics observed in the present suite of DNS.

\citet{daniel2023} investigated the transition to the Tayler-Spruit dynamo, but for negative rotation gradients, like in stellar radiative zones. In their setup, they trigger the stationary Tayler-Spruit dynamo via a first magnetic field amplification produced by a supercritical spherical Taylor-Couette dynamo. As already mentioned in our previous numerical studies~\citep{barrere2023,barrere2024a}, when $Ro<0$ the dynamo shows several significant differences from a physical point of view. For instance, the large-scale magnetic field harbours a quadrupolar symmetry and is mainly located in the equatorial plane (although it can be more localised within one hemisphere for weak stratification~\citep{petitdemange2024}). From a dynamical standpoint, \citet{daniel2023} model the subcritical transition to the Tayler-Spruit dynamo using a set of coupled amplitude equations. They manage to reproduce both the observed supercritical and subcritical transitions to the spherical Taylor-Couette dynamo and Tayler-Spruit dynamo, respectively. However, this modelling does not apply to our case, which shows more complex dynamics. Therefore, our study motivates further investigations of the observed difference between the $Ro>0$ and $Ro<0$ regimes in the spherical Taylor-Couette configuration to understand its origin. 

Finally, the diversity of dynamo regimes and temporal dynamics may have interesting consequences for the origin of NS magnetic fields. Recent studies suggest that the complex high-energy pulse profile of millisecond pulsars~\citep{bilous2019,petri2023,petri2025} and magnetars~\citep{guillot2015} can be explained by the presence of hot spots at their surface. Strong surface magnetic fields and magnetospheric currents can produce large variations in the temperature distribution. Therefore, predicting NS magnetic fields is essential to explain their high-energy emission. The Tayler-Spruit dynamo has been invoked to explain magnetar formation in a PNS spun up by fallback accretion, but the variety of dynamo regimes could reproduce the emission of other NS. On the one hand, numerical models of NS evolution suggest that the strong branch is a good candidate to explain the formation of magnetars with a ``weak'' magnetic dipole~\citep{igoshev2025}. On the other hand, the weaker magnetic field produced in the other states might explain the complex distribution of hot regions observed in millisecond pulsars, which require the presence of non-dipolar magnetic fields.~\citep{bilous2019,kalapo2021,petri2025}. Furthermore, the non-stationary dynamics of reversing states could produce different NS magnetic fields depending on the PNS magnetic field configuration when the dynamo stops. Thus, a possible extension of our study is to investigate the post-dynamo evolution of magnetic fields generated by these different states of the Tayler-Spruit dynamo in a realistic NS structure.

\begin{acknowledgments}
We thank the referees for their thorough reading and comments, which have been useful to improve the manuscript. The authors also thank A. Reboul-Salze for the fruitful discussions about the interpretation of the simulations. This work was supported by the European Research Council (MagBURST grant 715368) and the  PNPS and PNHE programs of CNRS/INSU, co-funded by CEA and CNES. Numerical simulations have been carried out at the CINES on the Jean-Zay supercomputer and at the TGCC on the supercomputer IRENE-ROME (DARI projects A0190410317 and A0170410317).

\end{acknowledgments}

\bibliography{biblio}
\end{document}